\documentclass[a4paper,11pt]{article}
\pdfoutput=1
\usepackage[normalem]{ulem}

\usepackage{jcappub} 
                     
\usepackage{amsmath}
\usepackage{amssymb}
\usepackage{color}
\usepackage{graphicx}
\usepackage{comment}
\usepackage{multirow}
\usepackage[dvipsnames]{xcolor}
\usepackage[utf8]{inputenc}
\usepackage{caption}

\usepackage[T1]{fontenc} 

\newcommand{\HII}{\rm H~{\sc ii }}

\title{Position-dependent power spectra of the 21-cm signal from the epoch of reionization}

\author[a]{Sambit K. Giri,}
\author[b]{Anson D'Aloisio,}
\author[a]{Garrelt Mellema,}
\author[c,d]{Eiichiro Komatsu,}
\author[a]{Raghunath Ghara}
\author[e,f]{and Suman Majumdar}

\affiliation[a]{Department of Astronomy and Oskar Klein Centre, \\Stockholm University, AlbaNova, SE-106 91 Stockholm, Sweden}
\affiliation[b]{Department of Physics \& Astronomy, \\ University of California, Riverside, CA 92521, USA}
\affiliation[c]{Max-Planck-Institute f$\mathrm{\ddot{u}}$r Astrophysik,\\Karl-Schwarzschild-Str. 1, Garching, 85741 Germany}
\affiliation[d]{Kavli Institute for the Physics and Mathematics of the Universe (Kavli IPMU, WPI),\\ Todai Institutes for Advanced Study, the University of Tokyo, Kashiwa, 277-8583 Japan}
\affiliation[e]{Centre of Astronomy,  Indian Institute of Technology Indore, Simrol, Indore 453552, India}
\affiliation[f]{Department of Physics, Blackett Laboratory, Imperial College, London SW7 2AZ, UK}
\emailAdd{sambit.giri@astro.su.se}
\emailAdd{anson.daloisio@gmail.com}
\emailAdd{garrelt.mellema@astro.su.se}
\emailAdd{komatsu@mpa-garching.mpg.de}
\emailAdd{raghunath.ghara@astro.su.se}
\emailAdd{suman.majumdar@iiti.ac.in}

\abstract{
The 21-cm signal from the epoch of reionization is  non-Gaussian.  Current radio telescopes are focused on detecting the 21-cm power spectrum, but in the future the Square Kilometre Array is anticipated to provide a first measurement of the bispectrum. Previous studies have shown that the position-dependent power spectrum is a simple and efficient way to probe the squeezed-limit bispectrum.  In this approach, the survey is divided into subvolumes and the correlation between the local power spectrum and the corresponding mean density of the subvolume is computed.  This correlation is equivalent to an integral of the bispectrum in the squeezed limit, but is much simpler to implement than the usual bispectrum estimators. It also has a clear physical interpretation: it describes how the small-scale power spectrum of tracers such as galaxies and the 21-cm signal respond to a large-scale environment. Reionization naturally couples large and small scales as ionizing radiation produced by galactic sources can travel up to tens of Megaparsecs through the intergalactic medium during this process. Here we apply the position-dependent power spectrum approach to fluctuations in the 21-cm background from reionization. We show that this statistic has a distinctive evolution in time that can be understood with a simple analytic model.
We also show that the statistic can easily distinguish between simple ``inside-out'' and ``outside-in'' models of reionization.  The position-dependent power spectrum is thus a promising method to validate the reionization signal and to extract higher-order information on this process.}

\keywords{reionization, non-gaussianity, power spectrum, first stars}

\begin{document}
\maketitle
\flushbottom

\section{Introduction}
\label{sec:intro}
The spin-flip transition of the ground state of neutral hydrogen corresponds to a rest frame wavelength of 21.1 cm, and is commonly known as the 21-cm signal. This signal has the potential to map the epoch of reionization (EoR) -- the last major phase change of the Universe when the intergalactic hydrogen transitioned from a cold and neutral state to a hot and ionized state \citep{1997ApJ...475..429M}. Recent indirect observations of the EoR suggest that reionization has completed at around a redshift of 6 \citep[e.g.][]{2015ApJ...802L..19R, 2015MNRAS.454L..76M, 2016A&A...594A..13P}; thus, we expect to find a 21-cm signal from the EoR at radio wavelengths longward of 1.47~m. Future 21-cm observations at different wavelengths will be able to follow the progress of the reionization process \cite[e.g.][]{2012RPPh...75h6901P}.

One of the challenging aspects of studying the EoR with the 21-cm signal is the optimal extraction of information. A new generation of radio telescopes, such as the Giant Metrewave Radio Telescope\footnote{\url{http://www.gmrt.tifr.res.in/}} \citep[GMRT; e.g.][]{2011MNRAS.413.1174P}, the Low Frequency Array\footnote{\url{http://www.lofar.org/}} \citep[LOFAR; e.g.][]{2010MNRAS.405.2492H}, the Murchison Widefield Array\footnote{\url{http://www.mwatelescope.org/}} \citep[MWA; e.g.][]{2009IEEEP..97.1497L} and the Precision Array for Probing the Epoch of Reionization\footnote{\url{http://eor.berkeley.edu/}} \citep[PAPER; e.g.][]{2010AJ....139.1468P} have been trying to detect the EoR statistically. However, they are only sensitive enough to measure the two point statistics (or power spectrum). Since the 21-cm signal from reionization is highly non-Gaussian \citep[e.g.][]{2014MNRAS.443.3090W,2016MNRAS.458.3003S}, the power spectrum will not provide a complete statistical description of this process. 

The Square Kilometre Array\footnote{\url{http://www.skatelescope.org/}} \citep[SKA;][]{2013ExA....36..235M} will be able to deliver images of the 21-cm background during the EoR. By combining images at different frequencies it will provide three-dimensional -- so-called tomographic -- data sets which will show the temporal evolution along the frequency direction. These observations will be sensitive to non-Gaussian information and thus will allow the first measurement of the 21-cm bispectrum -- the Fourier transform of the three-point correlation function. The bispectrum is non-zero for triangular configurations of three wave vectors in Fourier space, i.e., ${\mathbf k}_1+{\mathbf k}_2+{\mathbf k}_3=0$.  Since there are many different triangular configurations, it is customary to characterize the bispectrum in various limits for simplicity, such as equilateral ($k_1=k_2=k_2$), isosceles ($k_1>k_2=k_3$), and squeezed ($k_1\approx k_2\gg k_3$) triangles etc.

There are at least two hurdles in studying the 21-cm bispectrum. One is a computational challenge: it is computationally expensive to calculate all possible configurations of the bispectrum and their covariance matrix from the (simulated) observations \citep{2015MNRAS.451..467S, 2016MNRAS.458.3003S,2018MNRAS.476.4007M}, though there have been attempts to develop faster algorithms using the Fast Fourier Transform \citep[FFT;][]{2017MNRAS.472.2436W}. The other is physical interpretation: it is not easy to understand the information content of the 21-cm bispectrum, which is complicated further by many possible choices for triangles.
Refs. \citep{2016MNRAS.458.3003S,2018MNRAS.476.4007M} have taken appreciable steps forward in this regard. 

In the context of lower-redshift large-scale structure surveys, Refs. \cite{2014JCAP...05..048C,chiangthesis:2015} introduced a computationally inexpensive method to probe the bispectrum in the squeezed limit, in which the wave vectors form isosceles triangles with the unequal side much smaller than the other two ($k_3 \ll k_1\approx k_2$). The method relies on the estimation of the local (or position dependent) power spectra of the signal: one divides the survey volume into subvolumes, and  correlates power spectra and mean densities of the tracers (such as the galaxies and the 21-cm signal) locally measured within the subvolumes.  This method was successfully applied to the Baryon Oscillation Spectroscopic Survey\footnote{\url{http://www.sdss3.org/surveys/boss.php}} (BOSS) to measure non-Gaussianity due to gravitational evolution and non-linear galaxy bias \cite{2015JCAP...09..028C}. 

In this paper we apply the position-dependent power spectrum approach to brightness fluctuations in the 21-cm background from reionization.
This method is useful not only for its computational advantage, but also for physical interpretation. It describes how the small-scale power spectrum of tracers {\it responds} to the large-scale environment, i.e., long-wavelength fluctuations. It therefore naturally and intuitively captures mode-coupling of large- and small-scale fluctuations, which sheds light on how reionization proceeds.

The rest of this paper is organized as follows. In Section~\ref{sec:formalism} we summarize the position-dependent power spectrum approach of Refs.~\cite{2014JCAP...05..048C,chiangthesis:2015}.  As we do not have real 21-cm observations yet, we rely on simulated observations, which are described in Section~\ref{sec:21-cm_sig}. We show our main results in Section~\ref{sec:mainresults}, and show how the position-dependent power spectrum approach can distinguish between ``inside-out" and ``outside-in" reionization scenarios in Section~\ref{sec:source_models}. We conclude in Section~\ref{sec:discuss}. 
We explain our analytical model for the response function on large scales in Appendix~\ref{sec:toy_model}. 

\section{Formalism}
\label{sec:formalism}
\subsection{Redshifted 21-cm signal}
\label{sec:bright_temp}
The intensity of the 21-cm background can be expressed as a differential brightness temperature with respect to the Cosmic Microwave Background (CMB) \citep[e.g.,][]{2012RPPh...75h6901P,2013ExA....36..235M},
\begin{eqnarray}
T_{21} (\mathbf{r}, z) \approx 27~\mathrm{mK}~ x_\mathrm{HI}(\mathbf{r}) (1 + \delta_\mathrm{d}(\mathbf{r}))\left( \frac{1+z}{10} \right)^\frac{1}{2}
\left( 1 -\frac{T_\mathrm{CMB}(z)}{T_\mathrm{s}(\mathbf{r})} \right)\nonumber\\
\times
\left(\frac{\Omega_\mathrm{b}}{0.044}\frac{h}{0.7}\right)
\left(\frac{\Omega_\mathrm{m}}{0.27} \right)^{-\frac{1}{2}} 
\left(\frac{1-Y_\mathrm{p}}{1-0.248}\right)\,.
\label{eq:dTb}
\end{eqnarray}
In the above equation there are three position-dependent quantities: $x_\mathrm{HI}$ is the neutral fraction of hydrogen, $\delta_\mathrm{d}$ is the matter density contrast defined in the usual way, and $T_\mathrm{s}$ is the excitation temperature of the two spin states of the hydrogen ground state, known as the spin temperature. The global quantities are $T_\mathrm{CMB}(z)$, the CMB temperature at redshift $z$, $\Omega_\mathrm{b}$ and $\Omega_\mathrm{m}$, the present-day density parameters of baryons and total matter, and the primordial helium mass fraction, $Y_\mathrm{p}$.  

There will be no 21-cm signal when $T_\mathrm{CMB} =T_\mathrm{s}$. It is expected that in the early phases of reionization the spin temperature will decouple from $T_\mathrm{CMB}$ and approach the gas temperature due to the Wouthuysen-Field effect \citep{field1958excitation,wouthuysen1952excitation,1997ApJ...475..429M}.  Previous studies have shown that the inter-galactic medium (IGM) will likely be heated by the first X-ray sources before substantial reionization starts \citep{pritchard200721}. 
Therefore we will adopt the common assumption that the spin temperature is much greater than the CMB temperature, $\left(1-\frac{T_\mathrm{CMB}}{T_\mathrm{s}}\right) \rightarrow 1.$\footnote{It is expected that this assumption breaks down at higher redshifts, but exactly when that occurs is highly uncertain.}  Then, spatial variations in the 21-cm signal  depend only on the neutral fraction and density and eq.~(\ref{eq:dTb}) can be rewritten as
\begin{equation}
\label{eq:T21}
T_{21}(\mathbf{r}, z)=\hat{T}_\mathrm{21} x_\mathrm{HI}(\mathbf{r}) (1 + \delta_\mathrm{d}(\mathbf{r}))\,,
\end{equation}
where $\hat{T}_\mathrm{21}$ contains all the global cosmological terms in eq.~(\ref{eq:dTb}).

Observations will provide three-dimensional (tomographic) data sets for $T_{21}$ consisting of images at different frequencies. Along the frequency direction these data sets will be affected by various line-of-sight effects such as the light-cone effect \citep{2012MNRAS.424.1877D,giri2017bubble} and redshift space distortions \citep{2013MNRAS.435..460J,2016MNRAS.456...66J}. We will ignore these effects in this initial study.

\subsection{Position-dependent power spectrum as a probe of the squeezed-limit bispectrum}

In this section, we formulate the position-dependent power spectrum and its connection to the squeezed-limit bispectrum following Refs.~\cite{2014JCAP...05..048C,chiangthesis:2015}. Position-dependent power spectrum is an intuitive statistic that captures mode-coupling between large- and small-scale fluctuations. Instead of measuring the power spectrum from the entire survey volume, we divide it into many subvolumes, in which we compute the local power spectra and local mean density fields. The correlation between these two quantities measures coupling between the long-wavelength fluctuation (the local mean density field) and the short-wavelength fluctuation (the local power spectrum).

The correlation between the local matter density power spectra and the local mean matter density field is easy to understand. As the structure formation proceeds faster in overdense regions, the local power spectrum is larger than the average. Therefore there is a positive correlation between the local matter density power spectra and mean matter density field. We can understand this correlation by treating each subvolume as a ``separate universe'' \cite{1980lssu.book.....P}; namely, an overdense region behaves as if it was a separate Friedmann-Lema\^itre-Robertson-Walker universe with positive spatial curvature, even if geometry of the background universe is flat. This picture allows us to calculate theoretically the correlation between the local matter density power spectra and mean matter density field using perturbation theory \cite{2014JCAP...05..048C} in a mildly non-linear regime and the so-called separate universe simulation in a deeply non-linear regime \cite{2015MNRAS.448L..11W}.

While the position-dependent power spectrum measures only a fraction of information contained in the full bispectrum, the fact that we can understand it physically and intuitively is a unique advantage of this approach, as the bispectrum is often too complicated to understand physically.

We do not have to use the same tracer for the local power spectra and the local mean density field. For example, one can correlate the local Lyman-$\alpha$ forest power spectra with the large-scale fluctuation fields of quasars \cite{2017JCAP...06..022C} and weak lensing \cite{2016PhRvD..94j3506D,2018JCAP...01..012C}. Such correlations yield information on non-gravitational effects such as radiative transfer. In this paper we consider the correlation between the local 21-cm power spectra and the mean matter density field or the mean 21-cm field.

Let $\delta(\mathbf{r})$ be an arbitrary zero-mean field. (For example, below we will consider the matter density contrast, $\delta_\mathrm{d}$, as well as 21-cm brightness temperature fluctuations, $\delta_{21}=T_\mathrm{21}/\bar{T}_\mathrm{21}-1$.)   The position-dependent power spectrum of $\delta(\mathbf{r})$ can be calculated by dividing the total volume into subvolumes $V_L$.  Formally, the subvolumes are represented by a window function $W_L(\mathbf{r}-\mathbf{r}_L)$ centered on a position $\mathbf{r}_L$.  In what follows, we use the coordinate-space top-hat function, 

\begin{equation}
W(\mathbf{x}) = \prod_{i=1}^{3} \theta (x_i),~~~\theta (x_i) = \left\{\begin{matrix}
1, \ |x_i| \leq L/2,\\ 
0, \ |x_i| > L/2,
\end{matrix}\right.
\end{equation}
where $L$ is the length of each side of the subvolume.  The position-dependent power spectrum is then given by
 
\begin{equation}
P (\mathbf{k},\mathbf{r}_L) = \frac{1}{V_L}|\delta(\mathbf{k},\mathbf{r}_L)|^2\,,
\label{eq:PDS_def}
\end{equation}

\noindent where

\begin{eqnarray}
\nonumber
\delta (\mathbf{k},\mathbf{r}_L)  &=&\int \mathrm{d}^3r~\delta(\mathbf{r})W_L(\mathbf{r}-\mathbf{r}_L)e^{-i\mathbf{k}.\mathbf{r}}\\
&=& \int \frac{\mathrm{d}^3q}{(2\pi)^3}~\delta(\mathbf{k}-\mathbf{q})W_L(\mathbf{q})~e^{-i\mathbf{q}.\mathbf{r}_L}\,.
\label{local_field}
\end{eqnarray}
We find

\begin{equation}
P(\mathbf{k},\mathbf{r}_L)=\frac{1}{V_L}\int \frac{\mathrm{d}^3q_1}{(2\pi)^3} \int \frac{\mathrm{d}^3q_2}{(2\pi)^3}~\delta(\mathbf{k}-\mathbf{q}_1)\delta(-\mathbf{k}-\mathbf{q}_2)W_L(\mathbf{q}_1)W_L(\mathbf{q}_2)~e^{-i\mathbf{r}_L.(\mathbf{q}_1+\mathbf{q}_2)}.
\end{equation}

Defining $\bar{\delta}(\mathbf{r}_L)$ to be the mean of $\delta(\mathbf{r})$ in a subvolume $V_L$, the cross-correlation with $P(\mathbf{k},\mathbf{r}_L)$ can be written as
\begin{eqnarray}
\label{eq:int_bispec}
\left \langle P(\mathbf{k},\mathbf{r}_L) \bar{\delta}(\mathbf{r}_L) \right \rangle = \frac{1}{V_L^2} \int \frac{\mathrm{d}^3q_1}{(2\pi)^3} \int \frac{\mathrm{d}^3q_2}{(2\pi)^3} \int \frac{\mathrm{d}^3q_3}{(2\pi)^3} \left \langle \delta(\mathbf{k}-\mathbf{q}_1) \delta(-\mathbf{k}-\mathbf{q}_2) \delta(-\mathbf{q}_3) \right \rangle \nonumber\\
W_L(\mathbf{q}_1) W_L(\mathbf{q}_2) W_L(\mathbf{q}_3)~e^{-i\mathbf{r}_L.(\mathbf{q}_1+\mathbf{q}_2+\mathbf{q}_3)} \nonumber\\
= \frac{1}{V_L^2} \int \frac{\mathrm{d}^3q_1}{(2\pi)^3} \int \frac{\mathrm{d}^3q_3}{(2\pi)^3} B(\mathbf{k}-\mathbf{q}_1,-\mathbf{k}+\mathbf{q}_1+\mathbf{q}_3,-\mathbf{q}_3) \nonumber\\
W_L(\mathbf{q}_1) W_L(\mathbf{q}_1+\mathbf{q}_3) W_L(\mathbf{q}_3) \ .
\end{eqnarray}
Here, $B$ is the bispectrum, 

\begin{equation}
\left \langle \delta(\mathbf{q}_1) \delta(\mathbf{q}_2) \delta(\mathbf{q}_3) \right \rangle = B(\mathbf{q}_1,\mathbf{q}_2,\mathbf{q}_3) (2\pi)^3 \delta_\mathrm{D}(\mathbf{q}_1+\mathbf{q}_2+\mathbf{q}_3) \ ,
\end{equation}
where $\delta_\mathrm{D}$ is the Dirac delta function.  Following Ref.~\cite{2014JCAP...05..048C}, we define the right hand side of eq.~(\ref{eq:int_bispec}) to be the integrated bispectrum, denoted as $iB(\mathbf{k})$.  In this study, we will consider the spherically-averaged integrated bispectrum,

\begin{equation}
iB(k) \equiv  \left \langle P(k,\mathbf{r}_L) \bar{\delta}(\mathbf{r}_L) \right \rangle,
\label{eq:int_bispec_sphavg}
\end{equation}
where $P(k,\mathbf{r}_L)$ is the spherically averaged local power spectrum of the tracer within a subvolume centered at $\mathbf{r}_L$.  Since $W_L(\mathbf{q}_3)$ limits ${\mathbf q}_3$ to the scale defined by the size of subvolumes, $2\pi/L$,  the cross correlation, $\left\langle P(k,\mathbf{r}_L) \bar{\delta}(\mathbf{r}_L) \right \rangle$, probes the squeezed-limit bispectrum as long as $q_1$ and $q_2$ correspond to scales much smaller than our sub-volumes, i.e. $q_1 \approx q_1 \gg q_3$. When applied to the 21-cm signal, eq.~(\ref{eq:int_bispec_sphavg}) has the key virtue that it contains more information than the power spectrum, but is simpler to estimate compared to the conventional bispectrum estimators \citep{2016MNRAS.458.3003S,2017MNRAS.472.2436W,2018MNRAS.476.4007M}. 

For physical interpretation, it is useful to think of the local power spectrum as Taylor expansion in terms of $\bar\delta$, i.e.,
\begin{equation}
    P(k,\mathbf{r}_L) = P(k) + \left.\frac{dP(k,\mathbf{r}_L)}{d\bar{\delta}}\right|_{\bar\delta=0}\bar{\delta}+\cdots\,,
\end{equation}
where $P(k)$ is the global power spectrum. The second term describes how the local power spectrum responds to $\bar\delta$. To extract this information, let us define the so-called {\it response function} \cite{2014JCAP...05..048C,chiangthesis:2015} as
\begin{equation}
\label{eq:factorized_iBk}
f(k) \equiv \frac{iB(k)}{\sigma_L^2 P(k)},
\end{equation}
where $\sigma_L^2$ is the variance of $\bar{\delta}_L$.  To first order in $\bar\delta$, the response function reduces to
\begin{eqnarray}
\label{eq:resp_ln_form}
f(k) = \frac{d\mathrm{ln}P(k,\mathbf{r}_L)}{d\bar{\delta}}\Big|_{\bar{\delta}=0}.
\end{eqnarray}
The integrated bispectrum defined by eq.~(\ref{eq:int_bispec_sphavg}) thus measures the response of the local power spectrum to long-wavelength fluctuations. 
See Refs.~\cite{2014JCAP...05..048C,2015MNRAS.448L..11W,chiangthesis:2015} for the way to calculate $f(k)$ theoretically using perturbation theory and separate universe simulations, when both $P(k,\mathbf{r}_L)$ and $\bar\delta$ refer to the underlying matter density fields. In linear theory, $f_{\rm d,d}(k)=68/21-(1/3)d\ln [k^3P_{\rm dd}(k)]/d\ln k$, where $P_{\rm{dd}}(k)$ is the linear matter power spectrum.

In this paper, we go beyond the matter density response function and consider the 21-cm signal response function. $P(k)$ in the rest of this paper therefore denotes the 21-cm power spectrum.
In Appendix A, we develop a simple analytical model for the response function of the 21-cm signal from reionization.  The chief assumption of the model is that, on large enough scales, the neutral fraction of hydrogen, $x_{\mathrm{HI}}$, is a biased tracer of the underlying density field.  Under this assumption, we find that the response functions of the 21-cm power spectrum to large-scale density and 21-cm brightness temperature fluctuations are given respectively by

\begin{equation}
\label{eq:f_21d_final}
{f}_\mathrm{21,d}(k) = \frac{2(b_1+b_2)}{b_1+1}  + \frac{d \mathrm{ln} P_\mathrm{dd}(k,\mathbf{r}_L)}{d \bar{\delta}_\mathrm{d}} \Big|_{\bar{\delta}_\mathrm{d}=0} \,,
\end{equation}

\begin{equation}
\label{eq:f_2121_final}
{f}_\mathrm{21,21}(k)= \frac{{f}_\mathrm{21,d}(k)}{\bar{x}_\mathrm{HI}(1+b_1)}\,.
\end{equation}
Here, $b_1$ and $b_2$ are the first and second local bias parameters of the neutral fraction, defined in eq.~(\ref{eq:b_N}), $P_{\rm{dd}}(k)$ is the linear matter power spectrum, and $\bar{x}_{\rm{HI}}$ is the global neutral fraction.  We will show that these simple expressions capture remarkably well the main features of the response functions measured from our reionization simulation.

\subsection{Estimators}
The response functions can be estimated by dividing our simulation box into $N_\mathrm{cut}^3$ subvolumes, where $N_\mathrm{cut}$ is the number of cuts in each direction. 
The shortest side of the squeezed triangle is given by $q_3 = 2\pi N_\mathrm{cut}/L_\mathrm{B}$, where $L_\mathrm{B}=714$~Mpc is the box size of our simulation described in Section~\ref{sec:21-cm_sig}. We can study different squeezed-limit triangle configurations by changing the value of $N_\mathrm{cut}$.

We estimate the cross correlation of small scale 21-cm power spectra with the large scale density fluctuations using the following estimator
\begin{equation}
\label{eq:bispectrum_estimator}
\hat{iB}_{21,\mathrm{d}}(k) = \frac{1}{N_\mathrm{cut}^3} \sum_{i=1}^{N_\mathrm{cut}^3} P(k,\mathbf{r}_{L,i})~\bar{\delta}_\mathrm{d}(\mathbf{r}_{L,i}) \ .
\end{equation}
We then divide $\hat{iB}_{21,\mathrm{d}}(k)$ by the average  power spectrum of all subvolumes $\bar{P}(k)=\frac{1}{N_\mathrm{cut}^3} \sum_{i=1}^{N_\mathrm{cut}^3} P(k)$ and the average variance of all subvolumes $\bar{\sigma}_L^2=\frac{1}{N_\mathrm{cut}^3} \sum_{i=1}^{N_\mathrm{cut}^3}\bar{\delta}^2_\mathrm{d}(\mathbf{r}_{L,i})$. This quantity is known as the normalized integrated bispectrum and is the estimator for the response function defined in eq.~(\ref{eq:factorized_iBk}),
\begin{equation}
\label{eq:norm_int_bs_21d}
\hat{f}_{21,\mathrm{d}}(k) = \frac{\hat{iB}_{21,\mathrm{d}}}{\bar{P}(k)\bar{\sigma}_L^2}\,.
\end{equation}
The estimator for $\hat{f}_{21,21}(k)$ is found by replacing $\bar{\delta}_\mathrm{d}$ by $\bar{\delta}_{21}$ in eqs.~(\ref{eq:bispectrum_estimator}) and (\ref{eq:norm_int_bs_21d}).

\section{Simulation of Reionization}
\label{sec:21-cm_sig}

We use a radiative transfer simulation of reionization to measure the 21-cm response functions. The simulation is performed in two steps. The first step is to run an N-body simulation of the matter density field and collapsed structures (halos) in a cosmological volume. This simulation is performed with the code CUBEP$^3$M\footnote{\tt http://wiki.cita.utoronto.ca/mediawiki/index.php/CubePM} \citep{2013MNRAS.436..540H}. We used 6912$^3$ particles and the comoving length of each side of the volume is 714 Mpc. The cosmological parameters used here are $\Omega_m$=0.27, $\Omega_k$=0, $\Omega_b$=0.044, $h=0.7$, $n=0.96$ and $\sigma_8$=0.8, consistent with the \textit{Wilkinson Microwave Anisotropy Probe} (WMAP) \citep{2011ApJS..192...18K,hinshaw2013nine} and \textit{Planck} \citep{2016A&A...596A.108P,2018arXiv180706209P} results. 

We postprocess the matter density fields from the N-body simulation with C$^2$-RAY\footnote{\tt https://github.com/garrelt/C2-Ray3Dm} 
\citep{2006NewA...11..374M}, a fully numerical 3D radiative transfer code. Our radiative transfer calculations are performed on a uniform grid of size $N_{\rm{rt}}=600^3$. We populate the dark matter haloes from our N-body simulation with ionizing sources.  Here we assume that only haloes with masses above $10^9$~M$_\odot$ release ionizing photons into the IGM \citep{2016MNRAS.456.3011D}.
We take the rate of ionizing photon production to be proportional to the masses such that 3 ionizing photons per halo baryon escape every $10^7$~years. These assumptions result in a reionization history which is consistent with the existing observational constraints \citep[][]{2015ApJ...802L..19R, 2015MNRAS.454L..76M, 2016A&A...596A.108P}.  C$^2$RAY employs the short characteristics ray tracing method to solve the radiative transfer equation \citep{raga19993d,lim20033d}.  The ray-tracing is performed up to a comoving distance of 71~Mpc from each source. This limit is meant to approximate the effect of the presence of optically thick absorbers in the IGM that are unresolved in our N-body simulation. For more details we refer the reader to Refs.~\cite{2006MNRAS.372..679M, 2006MNRAS.369.1625I, 2012MNRAS.424.1877D}. 

We will use the radiative transfer simulation described above as our fiducial simulation, labelled as FN-600. 
For reference, the volume(mass)-weighted mean ionized fraction is $20\%$, $50\%$, and $80\%$ at $z=8.26(8.59)$, $7.57(7.72)$, and $7.22(7.28)$, respectively.  The volume-weighted ionized fraction is lower than the mass-weighted fraction throughout reionization. This indicates that reionization of FN-600 has an ``inside-out'' nature. The Thomson scattering optical depth for this reionization history is $\tau=0.056$, consistent with the \textit{Planck} results \citep[][]{2016A&A...596A.108P,2018arXiv180706209P}.  We note that there is evidence from high-$z$ Lyman-$\alpha$ forest observations that reionization may have ended somewhat later than $z=6$ (in contrast to the models considered here, in which reionization ends before $z=7$).  This would not affect our main conclusions, however; here we are focused on describing the general features of the 21-cm response functions.

\section{Results}
\label{sec:mainresults}
\subsection{Local mean fields}
\label{sec:PdPS}
Before investigating the position-dependent power spectrum and response functions, we  first show behaviour of the local mean fields, $\bar{\delta}_\mathrm{d}$ and $\bar\delta_{21}$.
Here and below we limit ourselves to the case $N_\mathrm{cut}=2$.

 \begin{figure}[t] 
 \centering
  \includegraphics[width=\textwidth]{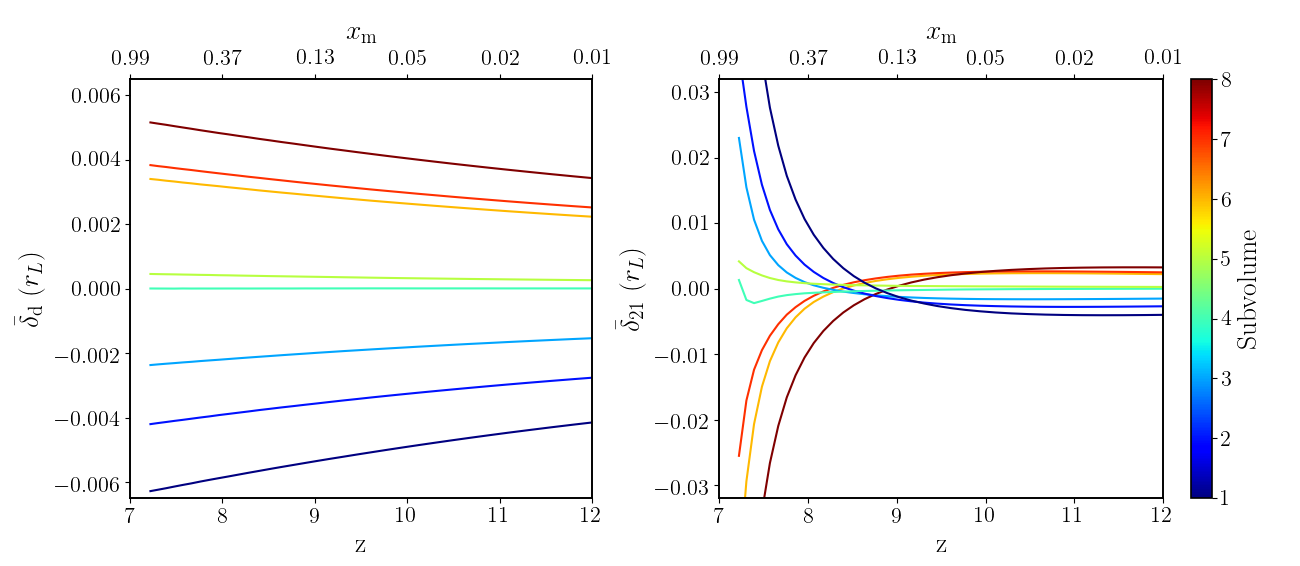}
 \caption{Redshift evolution of the long-wavelength mode, $\bar{\delta}_\mathrm{d}$ (left) and $\bar{\delta}_\mathrm{21}$ (right), in the FN-600 simulation. The comoving density of higher density subvolumes increases with time while the reverse is true for lower density subvolumes. The mean 21-cm signal shows the opposite behaviour. The signal in subvolumes with a lower $\bar{\delta}_\mathrm{21}$ at high $z$ increases with time and vice versa. This is indicative of the `inside-out' nature of the reionization.}
 	\label{fig:mean_subv}
 \end{figure}

Figure~\ref{fig:mean_subv} shows the redshift evolution of $\bar{\delta}_\mathrm{d}$ and $\bar\delta_{21}$ of the 8 subvolumes during reionization in our fiducial numerical simulation FN-600.  For reference, the top axis shows the mass-weighted mean ionized fraction ($x_\mathrm{m}$) in the simulation.  The left and right panels show $\bar{\delta}_\mathrm{d}$ and $\bar{\delta}_{21}$ for each subvolume, respectively. For ease of comparison, the colours representing each subvolume are matched between the left and right panels. We find that each subvolume evolves quite differently. For the mass density, the result is easy to understand: the denser subvolumes increase their density with time while the less dense ones decrease their density. The curves for different subvolumes never cross. 

The values of $\bar{\delta}_{21}$ on the other hand display more complex evolution. The behaviour of $\bar{\delta}_{21}$ is similar to that of the matter density field when the Universe is nearly neutral. However, since the denser regions reionize earlier in the ``inside-out'' scenario of reionization,  $\bar{\delta}_{21}$ of the densest region quickly starts decreasing (red curves).  By redshift 10, it is no longer the region with the highest $\bar{\delta}_{21}$; by redshift 8.8, it has the {\it lowest} $\bar{\delta}_{21}$. At around the same redshift the subvolume with the lowest density has the highest $\bar{\delta}_{21}$. 
The evolutionary curves for different subvolumes cross each other just after $z\approx 9$ ($x_m\approx 0.1$).

\subsection{Local 21-cm power spectrum vs matter density field}
\label{sec:PdPS_densityfield}

\begin{figure}[t]
 \centering
  \includegraphics[width=\textwidth]{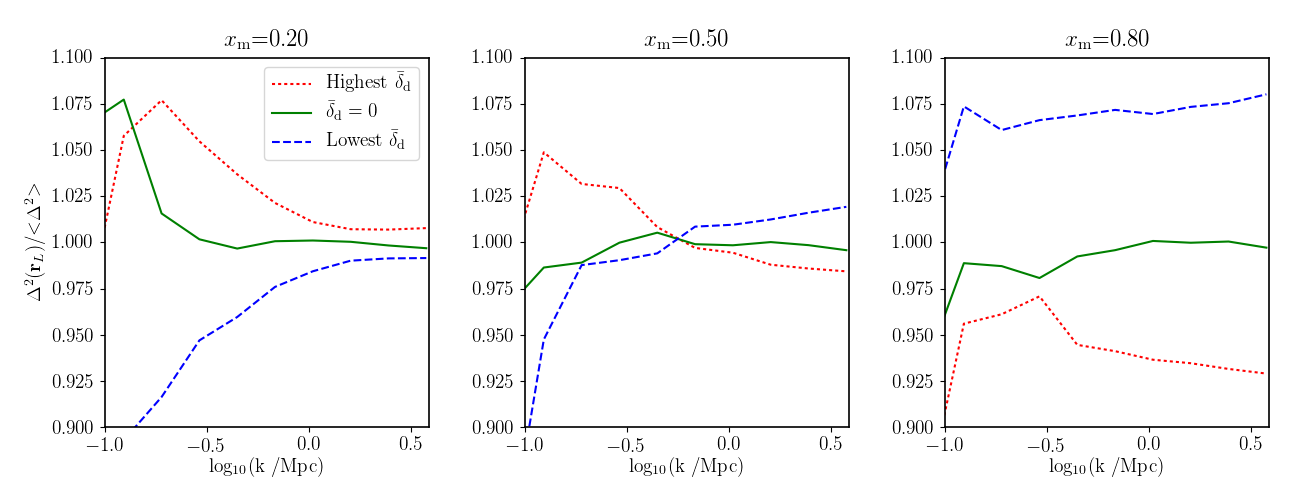}
 \caption{Redshift evolution of the position-dependent power spectra of the 21-cm signal constructed from the FN-600 simulation at three different reionization epochs. For clarity each of the spectra is normalised by the average of all the spectra at that epoch. In each panel, we show the spectra from the subvolume with the highest (red, dotted), mean (green, solid), and lowest (blue, dashed) $\bar{\delta}_\mathrm{d}$ at the corresponding epoch.
 The reionization progresses from the left to right panels, which is marked with mean ionization fraction, $x_\mathrm{m}$, of the full simulation box. We see that the correlation with the long-wavelength mode flips as reionization progresses.}
 \label{fig:PdPS_NumSim_d}
 \end{figure}

We calculate the position-dependent power spectra of the 21-cm signal using eq.~(\ref{eq:PDS_def}). In Figure~\ref{fig:PdPS_NumSim_d}, we show the position-dependent dimensionless power spectrum, $\Delta^2 (k, \mathbf{r}_L) = \frac{k^3\mathrm{P}(k, \mathbf{r}_L)}{2\pi^2}$, constructed from FN-600. We have chosen the subvolumes with  $\bar{\delta}_\mathrm{d}$ close to zero and those close to the two extreme ends. The $\Delta^2 (k, \mathbf{r}_L)$ at each epoch is normalised by the average of all the $\Delta^2 (k, \mathbf{r}_L)$ at that epoch. In each panel, we indicate the stage of reionization by the mass-weighted mean ionized fraction of the full simulation box. We show results for $x_\mathrm{m}=0.2$, 0.5 and 0.8 ($z=8.34,7.76$ and $7.305$).

The size of the subvolume determines the shortest wavenumber, $2\pi N_{\rm cut}/L_B=0.0176$~Mpc$^{-1}$. The squeezed-limit bispectrum is achieved when the wavenumbers of the local power spectra are much greater than this value. In our choice of parameters, we find that the squeezed-limit bispectrum is achieved reasonably well for $k\gtrsim 0.5$~Mpc$^{-1}$ ($\log_{10}(k)\gtrsim -0.3$). Our description below will focus on this large $k$ regime, unless stated otherwise.

During the early stages of reionization, the 21-cm signal follows the matter distribution. Therefore, we see a positive correlation between the local 21-cm power and $\bar{\delta}_\mathrm{d}$, similar to what has been previously noted for the matter power spectrum \citep[see figure~1 in Ref.][]{2014JCAP...05..048C}.

However, once a non-negligible fraction of the simulation volume has been ionized, an anti-correlation develops at larger $k$ values. This is easy to understand.
The ionizing sources form \HII regions around them, which grow and overlap with time \citep{2004ApJ...613....1F,2012RPPh...75h6901P}. A characteristic scale of the \HII regions increases as reionization progresses \citep[e.g.][]{2004ApJ...613....1F,furlanetto2006characteristic,2007ApJ...669..663M,giri2017bubble}. This characteristic scale leaves an imprint on the 21-cm power spectra which appears as a ``knee''-shaped feature \citep{zaldarriaga200421,wyithe2004characteristic}. The 21-cm power at scales smaller than this feature is lowered by reionization\footnote{In figure~1 of Ref.~\citep{2008ApJ...680..962L}, we see that $\Delta^2$ changes the slope as the ionization fraction increases. The slope initially increases (up to $x_\mathrm{m}\lesssim 0.15$) but the slope starts decreasing when the knee feature develops. After $x_\mathrm{m}\approx 0.70$ the slope no longer changes but the amplitude of $\Delta^2$ decreases as more hydrogen is ionized. Each subvolume follows a different reionization history (seen in Figure~\ref{fig:mean_subv}) and the subvolumes which start reionizing earlier will be ahead in this slope changing process.}. As reionization progresses inside-out in our FN-600 simulation, the high-density subvolumes are ionized first, which suppresses the 21-cm power spectra earlier. This explains an anti-correlation between the 21-cm power spectrum at large $k$ and $\bar\delta_{\rm d}$. The negative correlation spreads to lower $k$ values as more volume is reionized. By $x_\mathrm{m}=0.8$, the local 21-cm power spectra are anti-correlated with $\bar{\delta}_\mathrm{d}$ at all wavenumbers.

\subsection{Local 21-cm power spectrum vs 21-cm brightness field}
\label{sec:PdPS_21cmfield}

While dependence of the local 21-cm power spectra on the local mean matter density field is useful for physical interpretation, it may not be immediately observable because we may not have an adequate tracer of the mean matter density field during the EoR. Therefore we investigate the dependence of the local 21-cm power spectra on the local mean 21-cm brightness field, which is readily observable using the same data set.

 \begin{figure}[t] 
 \centering
  \includegraphics[width=\textwidth]{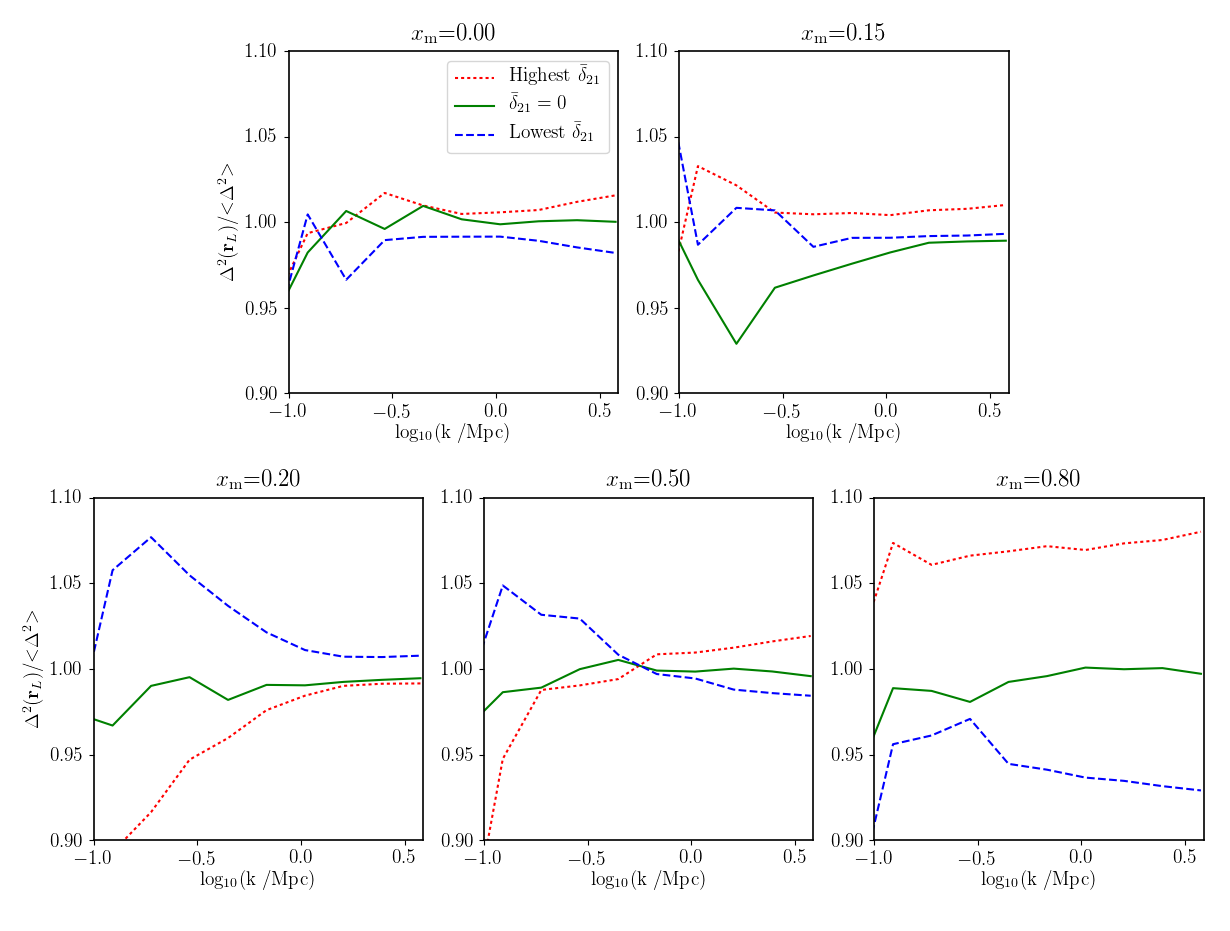}
 \caption{
 Same as Figure~\ref{fig:PdPS_NumSim_d} but with respect to $\bar\delta_{21}$ instead of $\bar\delta_{\rm d}$.
 }
 	\label{fig:PdPS_NumSim_21}
 \end{figure}

In Figure~\ref{fig:PdPS_NumSim_21}, we show $\Delta^2 (k, \mathbf{r}_L)$ for three values of $\bar\delta_{21}$. Interpretation is straightforward. At the very start when $x_\mathrm{m}=0$, the correlation with $\bar\delta_{21}$ is positive, as expected for the matter density power spectra. As reionization proceeds, we find the results that are opposite of those in Figure~\ref{fig:PdPS_NumSim_d}. Namely, the sign of the correlation flips because $\bar\delta_{\rm d}$ and $\bar\delta_{21}$ are anti-correlated for $x_\mathrm{m}\gtrsim 0.2$ (see Figure~\ref{fig:mean_subv}).
 The correlation of the local 21-cm power spectra with $\bar{\delta}_{21}$ therefore displays an additional phenomenon during the earliest phases of reionization: it evolves from a positive correlation, through an almost uncorrelated phase, to the anti-correlation seen at $x_\mathrm{m}= 0.2$. After this, the high $k$ power  shows a positive correlation due to the size of \HII regions when  $x_\mathrm{m} =0.5$. Eventually we find a positive correlation at all wavenumbers when  $x_\mathrm{m}=0.8$.

\subsection{21cm response functions}
\label{sec:resp_func}

We now turn our attention to the 21-cm response functions. 
Let us first consider the response with respect to the mean matter density field, $\hat{f}_{21,\mathrm{d}}$, which is more straightforward to interpret physically. We will then consider $\hat{f}_{21,21}$, which is the quantity that can be measured from observations.

\subsubsection{The matter-21cm-21cm response function}

\begin{figure}[t] 
 \centering
  \includegraphics[width=\textwidth]{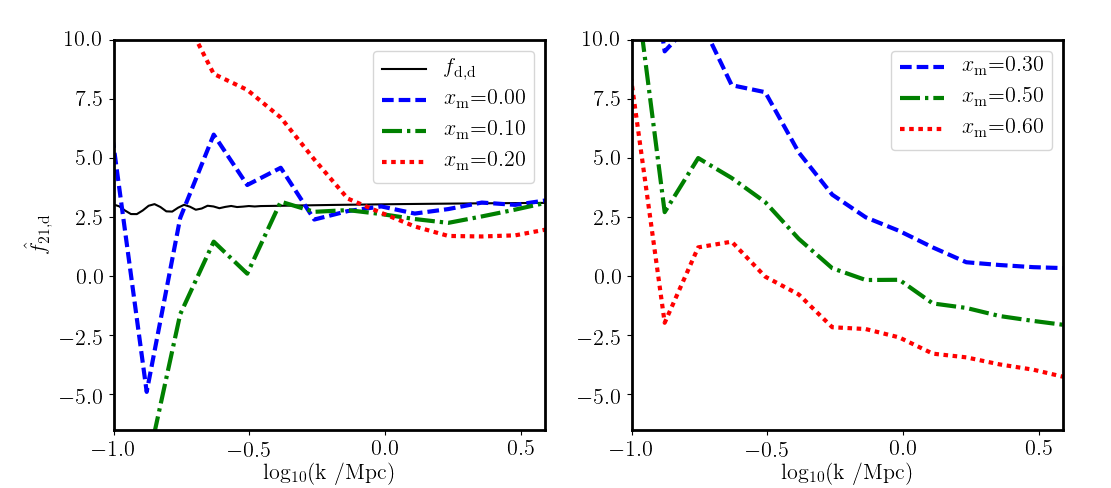}
 \caption{Response function of the local 21-cm power spectrum to the long-wavelength matter density field at different stages of reionization. We show the early and late stages in the left and right panels, respectively. In the beginning of reionization, the response function resembles the response function of the matter power spectrum. The squeezed limit is achieved for $\log_{10}(k)\gtrsim -0.3$. The black solid line shows the analytical response function of the linear matter power spectrum, $68/21-(1/3)d\ln [k^3P_{\rm dd}(k)]/d\ln k$.
 }
 	\label{fig:response_d2121}
 \end{figure}

Figure~\ref{fig:response_d2121} shows $\hat{f}_{21,\mathrm{d}}$ at different phases of reionization. The left panel shows the early phases ($x_\mathrm{m}=0.0$, 0.1 and 0.2) while the right panel shows the later ones ($x_\mathrm{m}=0.3$, 0.5 and 0.6). By construction, for $x_\mathrm{m}=0.0$ the 21-cm signal follows the density field exactly and thus the response function is identical to that of the matter density field. In our choice of parameters, the squeezed limit is achieved for $k\gtrsim 0.5$~Mpc$^{-1}$ ($\log_{10}(k)\gtrsim -0.3$), in which the response function agrees with $68/21-(1/3)d\ln [k^3P_{\rm dd}(k)]/d\ln k$ shown by the solid black line.

\begin{figure}[t]
 \centering
 $\renewcommand{\arraystretch}{-0.75}
   \begin{array}{c}
    \includegraphics[width=\textwidth]{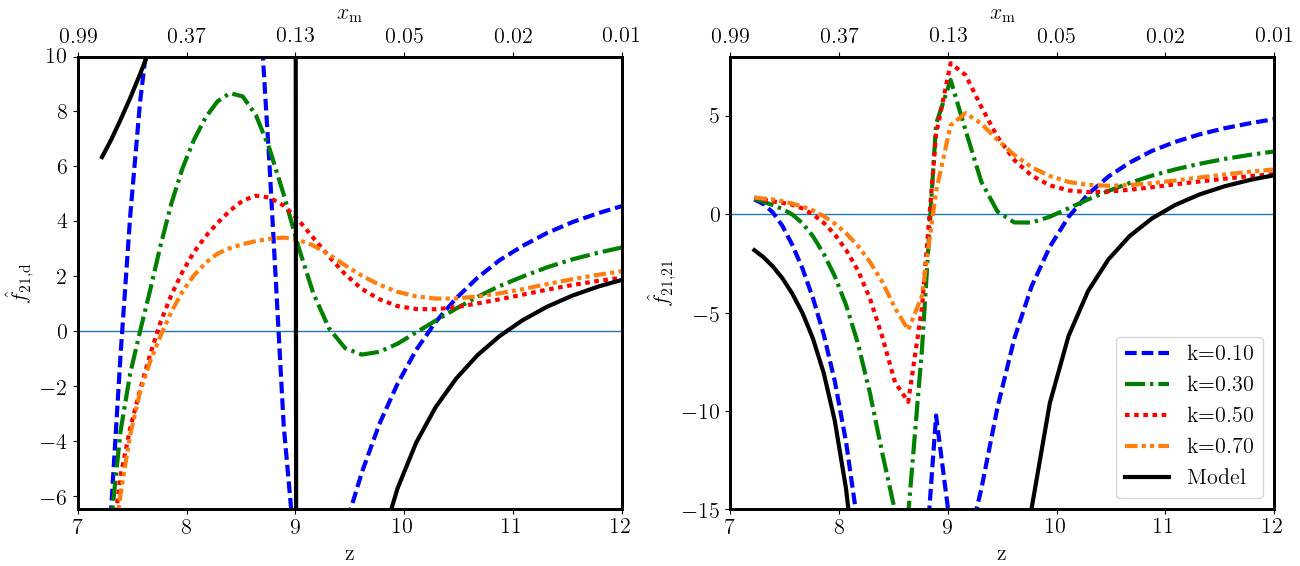}
  \end{array}$
 \caption{
 Redshift evolution of the response functions at various wavenumbers. The 21cm-21cm-matter and 21cm-21cm-21cm response functions are shown in the left and right panels, respectively. We see the sign change at the epochs predicted by our analytical model (solid black lines). 
 }
 \label{fig:response_vs_z}
 \end{figure}

At $x_\mathrm{m}=0.2$, the small-scale ($k\gtrsim 1$~Mpc$^{-1}$) response function decreases because of the size of \HII regions, while it still remains positive. At $x_\mathrm{m}=0.5$ the small-scale response shows a negative value, in agreement with the physical picture we gave in Section~\ref{sec:PdPS_densityfield}. The negative response then spreads to lower $k$ values as reionization proceeds. 

The behaviour of the response function on larger scales, $k\lesssim 0.5~{\rm Mpc}^{-1}$, is more complex, but we can reproduce it qualitatively using a simple analytical model we develop in Appendix A. In the left panel of Figure~\ref{fig:response_vs_z}, we compare the  redshift evolution of $f_\mathrm{21,d}$ for four different wavenumbers, $k=0.1$, 0.3, 0.5 and 0.7~Mpc$^{-1}$, with the prediction from the analytical model given in eq.~(\ref{eq:f_21d_final}). For $k\lesssim 0.5~{\rm Mpc}^{-1}$, the $f_\mathrm{21,d}$ decreases during the early phases of reionization, but then rapidly changes signs at $x_m \approx 0.13$. The timing of this sign-change can be understood using our model.  The crossover occurs when the linear bias of the neutral hydrogen with respect to the underlying density field is $b_1 = -1$.  At this time, the fluctuations in the hydrogen density are perfectly anti-correlated with the density fluctuations and the 21-cm power spectrum (to linear order) vanishes\footnote{In reality, higher order terms prevent the 21-cm power spectrum from vanishing.}.  In all of our models, this epoch of the minimum 21-cm power occurs at $x_m \approx 0.13$.  

While the model predicts a positive response function near the end of reionization, the simulation shows a negative correlation. This is because the size of \HII regions becomes larger than the spatial scales of the wavenumbers shown in this figure near the end of reionization, and our analytical model breaks down. However we have already understood the origin of the negative correlation: it is due to suppression of the 21-cm power spectrum by ionization.

\subsubsection{The 21cm-21cm-21cm response function}

\begin{figure}[t] 
 \centering
  \includegraphics[width=\textwidth]{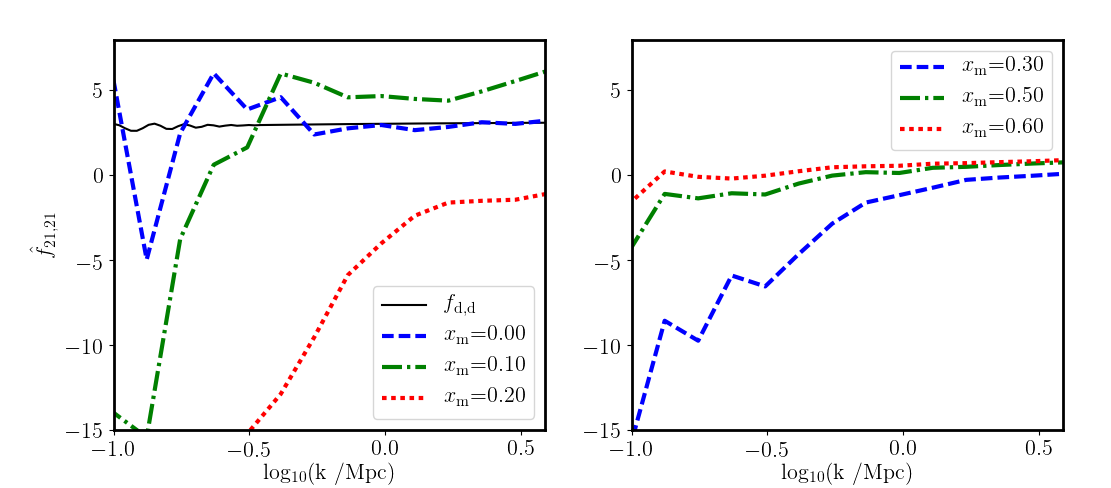}
 \caption{
 Same as figure~\ref{fig:response_d2121} but for the response function of the local 21-cm power spectrum to the long-wavelength 21-cm brightness field.
 }
 \label{fig:response_212121}
\end{figure}

While $\hat{f}_{21,\mathrm{d}}$ is simpler to interpret, we cannot construct it from the 21-cm observations alone. In this section we consider $\hat{f}_{21,21}$. Figure~\ref{fig:response_212121} shows $\hat{f}_{21,21}$ at the same global ionized fractions as in Figure~\ref{fig:response_d2121}.  

The signs of $\hat{f}_{21,\mathrm{d}}$ and $\hat{f}_{21,21}$ are opposite when $\bar\delta_\mathrm{d}$ and $\bar\delta_{21}$ become anti-correlated (Figure~\ref{fig:mean_subv}).
According to our model, $f_\mathrm{21,d}$ and $f_{21,21}$ only differ by a multiplication factor $[\bar{x}_\mathrm{HI}(1+b_1)]^{-1}$, see eq.~(\ref{eq:f_2121_final}). When $b_1 < -1$ the multiplication factor becomes negative and the model predicts that features in $f_\mathrm{21,21}$ will be inverted compared to those seen in $f_\mathrm{21,d}$. This is indeed seen in the simulation results. 
Unlike $\hat{f}_{21,\mathrm{d}}$, however, $\hat{f}_{21,21}$ rapidly approaches to zero as reionization proceeds for $x_\mathrm{m}\gtrsim 0.3$ (see the right panels of Figure~\ref{fig:response_vs_z} and \ref{fig:response_212121}).

\section{Inside-out versus outside-in reionization}
\label{sec:source_models}

 One of the most basic questions about the reionization process is whether it proceeded generally from high to low densities (``inside-out'') or from low to high densities (``outside-in'').   Answering this question would inform us about the nature of sources and sinks of ionizing photons \cite{2000ApJ...530....1M,2009MNRAS.394..960C}.
 Here we explore the 21-cm response functions as a powerful discriminant of these two scenarios.  
 
  \subsection{Models}
In addition to FN-600, we use two simple models to explore whether the 21-cm response function can discriminate between the inside-out and outside-in scenarios.  These models assume that the ionization state of a cell is fully determined by its local density. The only two inputs for these simulations are the mass-weighted mean ionized fraction, $x_\mathrm{m}(z)$, and the matter density field.  For the inside-out simulation (SN-in-out) the densest fraction $x_\mathrm{m}(z)$ of the mass is assumed to be fully ionized, whereas for the outside-in simulation (SN-out-in) the lowest density fraction $x_\mathrm{m}(z)$ is ionized. These two simulations therefore correspond to a perfect correlation and anti-correlation between ionization and density, respectively.  We note that these models are unrealistically extreme, but our purpose here is to demonstrate different behaviours of the response functions in the total inside-out and outside-in limits.  We use the same gridded density fields of $600^3$ for these models and use a reionization history (by mass) which approximately matches that of our radiative transfer simulation FN-600. Due to their similar reionization histories, the Thomson scattering optical depth is approximately the same for all models, $\tau\approx0.056$.

\subsection{Results}

\begin{figure}[t]
 \centering
  \includegraphics[width=\textwidth]{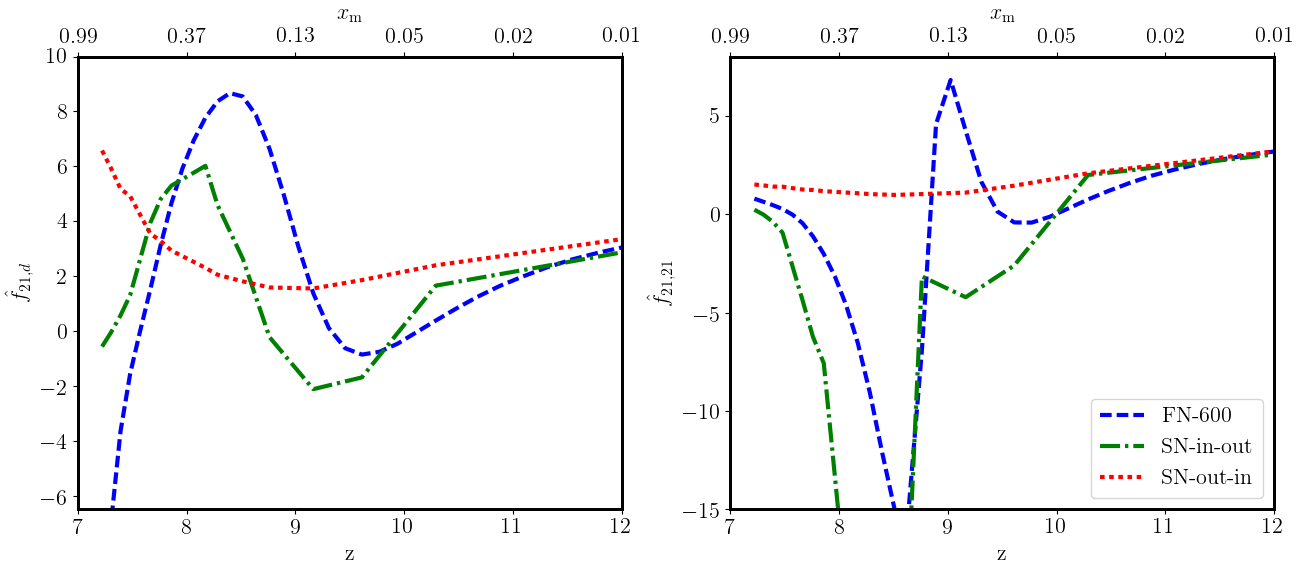}
 \caption{Redshift evolution of the response functions at $k=0.3$ Mpc$^{-1}$ for the three different reionization simulations. The 21cm-21cm-matter and 21cm-21cm-21cm response functions are shown in the left and right panels, respectively.}
 \label{fig:source_response_vs_z}
 \end{figure}

In Figure~\ref{fig:source_response_vs_z}, we compare the redshift evolution of the two response functions $\hat{f}_{21,\rm d}$ and $\hat{f}_{21,21}$ at $k=0.3~\mathrm{Mpc}^{-1}$ against the results from FN-600, SN-in-out, and SN-out-in. At $k=0.3~\mathrm{Mpc}^{-1}$, the error on the response function $\hat{f}_{21,21}$ will be minor ($\leq 0.1$) for 1000 h observation with SKA-Low. We provide a back-of-the-envelope calculation of the error on the $\hat{f}_{21,21}$ constructed from radio observations in Appendix~\ref{sec:back-of-envelope}.
The evolution in  SN-in-out is qualitatively similar to that of FN-600. We note, however, that  SN-in-out  shows a much deeper minimum for $\hat{f}_{21,21}$. 

The results of SN-out-in, on the other hand, show a very different evolution; $\hat{f}_{21,\rm d}$ always remains positive and increases throughout much of the reionization process. Unlike SN-in-out, $\hat{f}_{21,\rm d}$ does not display any extrema, and $f_\mathrm{21,21}$ is found to be almost flat throughout reionization.
In this scenario, the regions with higher densities remain unaffected until the very end of reionization. Thus, the bias of the neutral hydrogen with respect to the density filed ($b_1$ in the analytical model for the response function) will remain positive, and eqs.~(\ref{eq:f_21d_final}) and (\ref{eq:f_2121_final}) indicate that the response functions will never attain negative values. 
 
  \begin{figure}[t]
 \centering
  \includegraphics[width=\textwidth]{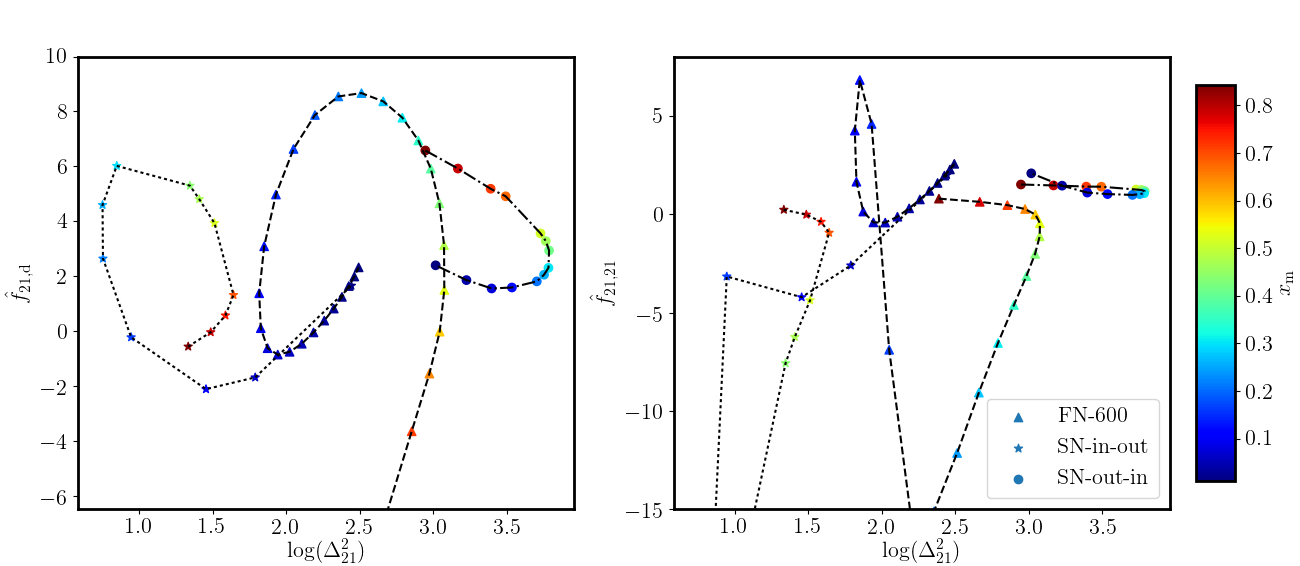}
 \caption{Evolution of reionization in the $f(k)-\Delta^2(k)$ plane at $k=0.3$ Mpc$^{-1}$ for the three different reionization simulations. The $f_\mathrm{21,d}$ and $f_\mathrm{21,21}$ response functions are shown in the left and right panels, respectively. The reionization epoch is given with the colour of the markers. 
  }
 \label{fig:pk_vs_bk}
 \end{figure}
 
To further illustrate how the response functions help discriminate between SN-in-out and SN-out-in,  we show in Figure~\ref{fig:pk_vs_bk} the evolutionary track of each simulation in the $f(k)-\Delta^2(k)$ plane for $k=0.3$ Mpc$^{-1}$. Ref.~\citep{2018MNRAS.476.4007M} introduced this representation in the study of their bispectrum results. The left panel shows the evolutionary tracks of $f_\mathrm{21,d}$ for FN-600, SN-in-out, and SN-out-in.  All the tracks start at roughly the same region of the $f(k)-\Delta^2(k)$ plane.  For the $f_\mathrm{21,d}$ case,  SN-in-out (SN-out-in)  move leftward (rightward) in the plane. Both SN-in-out and FN-600 create clockwise spiral tracks due to the simultaneous action of dip in the 21-cm power spectra and the sign change in $f_\mathrm{21,d}$. SN-out-in does not show this sign change in $f_\mathrm{21,d}$. Its evolutionary track is more parabolic in nature.  
The right panel of Figure~\ref{fig:pk_vs_bk} shows the evolutionary tracks of $f_\mathrm{21,21}$. Again, the tracks start at roughly the same region. Both the SN-in-out and FN-600 tracks exhibit a dip feature instead of a spiral owing to the fact that  $f_\mathrm{21,21}$ does not change the sign. The track of SN-out-in is similar to that of the left panel. These results show that the evolution of the 21-cm response function will provide a useful tool for probing the ``topology" of reionization.

\section{Summary and discussion}
\label{sec:discuss}

Due to non-Gaussian nature of the signal, higher order statistics such as the bispectrum are needed to extract the full information content of the 21-cm brightness fluctuations. In this paper we have investigated a particular choice for the bispectrum; namely, the squeezed-limit bispectrum. Following Refs.~\cite{2014JCAP...05..048C,chiangthesis:2015}, we have measured this bispectrum from the correlation between position-dependent 21-cm power spectra and large-scale variations in the matter density or the 21-cm signal fields. This statistic provides a clean measurement of mode-coupling of the large- and small-scale fluctuations, by capturing how the small-scale power in the 21-cm signal responds to the large-scale fluctuation field. Not only is this statistic easy to measure from the data, but it also provides a clear physical interpretation of the measured correlation. This property allowed us to derive a simple analytical model based on bias expansion and the excursion-set formalism. 

As structure formation in an overdense region proceeds faster than the average, we find a positive correlation between the small-scale matter power spectrum and the large-scale matter density field. In other words, the small-scale matter power has a positive response to the large-scale matter density field. When the dense regions are ionized first, as in the ``inside-out'' scenario, the response function of the small-scale 21-cm power spectrum is suppressed relative to the matter response function (e.g., at $k\gtrsim 1~{\rm Mpc}^{-1}$ for $x_m=0.2$), and becomes negative as reionization progresses (e.g., at $k\gtrsim 0.5~{\rm Mpc}^{-1}$ for $x_m=0.5$).  This negative response is seen in the scales below the typical sizes of \HII regions during reionization. 
Eventually we find a negative response at all scales towards the end of reionization.
In this way, reionization produces  mode-coupling naturally; large-scale \HII regions suppress the 21-cm power spectrum on small scales. The sign of the response of the small-scale 21-cm power spectrum to the large-scale 21-cm brightness field is opposite to this once the large-scale matter density and 21-cm brightness fields become anti-correlated.

While the behaviour of the response function below the size of \HII regions is understood as described above, the behaviour on larger scales ($k\lesssim 0.5~{\rm Mpc}^{-1}$) is more complex. To this end we have developed an analytical model that is valid on scales larger than the size of \HII regions.  The analytical model predicts that the detailed evolution of the response functions must depend on the distribution of neutral material with respect to the density field and hence on the distribution, brightness and clustering of sources of ionizing photons. We found that our model is able to explain the basic features of the response functions observed in our reionization simulation qualitatively.

 We applied the position-dependent power spectrum approach to two extreme models: one in which ionization correlates perfectly with density (perfect ``inside-out'' reionization) and one in which it anti-correlates perfectly (perfect ``outside-in'' reionization). We found that the response functions in these scenarios are very different, suggesting that the response functions are useful for studying the topology of reionization.  The response functions will also be useful to confirm that the detected 21-cm power spectra are not due to foreground emission or instrumental systematic errors. The power spectrum is susceptible to foreground and systematics because it is easy to add power at a given scale by these effects. However, the response function measures mode-coupling between large- and small-scale fluctuations, and reionization predicts a particular form of coupling which would be harder to mimic precisely by foreground or systematics. The unique evolution of the response function in the $f(k)-\Delta^2(k)$ plane can be used to establish reliability of the measurements. 

Our study only represents a first exploration of the application of the position-dependent power spectra and response function techniques to the 21-cm signal. A wider exploration of the parameters space of sources is needed. Another potentially important effect is the impact of spin temperature fluctuations. While we assumed the high spin temperature limit, the spin temperature may be closer to the CMB temperature  during an early phase of reionization, 
which will impact the strength of the 21-cm signal. Refs.~\citep{2017MNRAS.468.3785R,2018arXiv180803287R} showed that spin temperature fluctuations can have a strong impact on non-Gaussianity of the signal. Ref.~\citep{2018arXiv180802372W} indeed confirmed that there is a clear effect present in the measurements of the bispectrum.

How about observational prospects of the response functions? Addressing this question requires dividing the observational data set into subvolumes and extracting the local power spectrum and mean signal from each of these. Implementation of this procedure on real interferometeric data should be studied, not only from the point of view of the signal-to-noise ratio but also from that of calibration effects. Ref.~\citep{2018arXiv180802372W} have carried out a theoretical study of detectability of the bispectrum in SKA-Low observations in the presence of telescope noise. They find that for 1000-hour integration time the bispectrum from the equilateral triangle configuration will be detectable. A similar study should be done for the response function, hence the squeezed-limit bispectrum.

The squeezed limit bispectrum (as measured through position-dependent power spectrum method) clearly constitutes a powerful probe of reionization and should be added to the palette of analysis methods for the 21-cm signal from the EoR. The arrival of real observational data, in hopefully the not-too-distant future, will establish which of the many available methods are the most useful for confirming the nature of the detected signal as well as for extraction of astrophysical and cosmological parameters from it. The results of our initial study give us confidence that just as for galaxy surveys at lower redshifts \cite{2015JCAP...09..028C}, the squeezed-limit bispectrum will prove to be a valuable tool.

\acknowledgments

This work was supported by Swedish Research Council grant 2016-03581. We acknowledge that the results in this paper have been achieved using the PRACE Research Infrastructure resources Curie based at the Très Grand Centre de Calcul (TGCC) operated by CEA near Paris, France and Marenostrum based in the Barcelona Supercomputing Center, Spain. Time on these resources was awarded by PRACE under PRACE4LOFAR grants 2012061089 and 2014102339 as well as under the Multi-scale Reionization grants 2014102281 and 2015122822. Some of the numerical computations were done on the resources provided by the Swedish National Infrastructure for Computing (SNIC) at PDC, Royal Institute of Technology, Stockholm. 
SM acknowledges financial support from the European Research Council under ERC grant number 638743-FIRSTDAWN. EK was supported in part by JSPS KAKENHI Grant Number JP15H05896.

\appendix
\section{Analytical model for response function of the 21-cm power spectrum on large scales}
\label{sec:toy_model}
In this appendix we develop a simple analytic model for the 21-cm response functions ${f}_\mathrm{21,d}$ and ${f}_\mathrm{21,21}$. This model is based on a  bias expansion and the excursion set model of reionization \cite{2004ApJ...613....1F}. 

We start with eq.~(\ref{eq:T21}) which gives the 21-cm differential brightness temperature at some location $\mathbf{r}$ and 
define the neutral fraction fluctuation $\delta_\mathrm{HI}(\mathbf{r}) = x_\mathrm{HI}/\bar{x}_\mathrm{HI}-1$.
We assume that the scales of interesting are large enough to apply linear perturbation theory.  Indeed Ref.~\citep{2018arXiv180608372M} found that the 21-cm signal is likely perturbative over most of the scales accessible to forthcoming observations.  Expressing eq.~(\ref{eq:T21}) in terms of $\delta_\mathrm{HI}$, and keeping only terms linear in $\delta_\mathrm{d}$ and $\delta_\mathrm{HI}$ yields
\begin{equation}
\label{eq:dT21}
\Delta T_\mathrm{21}(k) = \hat{T}_\mathrm{21} \bar{x}_\mathrm{HI} (\delta_\mathrm{HI}(\mathbf{k})+\delta_\mathrm{d}(\mathbf{k}))\,,
\end{equation}
where we consider the 21-cm fluctuation, $\Delta T_\mathrm{21}=T_\mathrm{21}-\bar{T}_\mathrm{21}$, in Fourier space.

Assuming that we are interested in scales that are much larger than the typical ionized
bubble scale\footnote{Therefore, our analytical model cannot reproduce the response function in the small scales below the bubble size.}, we can express $\delta_\mathrm{HI}$ in terms of a linear bias parameter, $\delta_\mathrm{HI} (\mathbf{k})=b_1\delta_\mathrm{d}(\mathbf{k})$. To
first order, the 21-cm power spectrum can then be written as
\begin{equation}
\label{eq:P21}
P(k)=\hat{T}^2_\mathrm{21}\bar{x}^2_\mathrm{HI}(b_1+1)^2 P_\mathrm{dd}(k) \,,
\end{equation}
where $P_\mathrm{dd}(k)$ is the matter power spectrum. We compute the 21-cm response function by
taking the derivative of eq.~(\ref{eq:P21}) with respect to a long-wavelength density perturbation $\delta_\mathrm{d}^l$ and expressing it as eq.~(\ref{eq:resp_ln_form}),
\begin{equation}
\label{eq:f_21d}
\frac{\mathrm{d}  \mathrm{ln} P(k)}{\mathrm{d} \delta_\mathrm{d}^l} \Big|_{\delta_\mathrm{d}^l=0}= \left[ 2 b_1 +\frac{2}{b_1+1} \frac{\mathrm{d}  b_1}{\mathrm{d}  \delta_\mathrm{d}^l} \right]_{\delta_\mathrm{d}^l=0} + \frac{\mathrm{d}  \mathrm{ln} P_\mathrm{dd}(k)}{\mathrm{d}  \delta_\mathrm{d}^l} \Big|_{\delta_\mathrm{d}^l=0} \, .
\end{equation}
The last term on the right hand side is the response of the matter power
spectrum $f_\mathrm{d,d}$. To evaluate eq.~(\ref{eq:f_21d}) we just need a model for the bias $b_1$ and its derivative $\mathrm{d} b_1/\mathrm{d} \delta_\mathrm{d}^l$.

We obtain these bias parameters by applying the excursion-set model of reionization. By analogy to the excursion-set model of halo formation, let us define the local bias parameters of the neutral fraction to be
\begin{equation}
\label{eq:b_N}
b_N \equiv \frac{1}{\bar{x}_\mathrm{HI}} \frac{\mathrm{d} ^N x_\mathrm{HI}}{\mathrm{d}  (\delta_\mathrm{d}^l)^N} \Big|_{\delta_\mathrm{d}^l=0} \, .
\end{equation}
In the excursion-set model of reionization, the mean neutral fraction in some region of scale
$l$, with mean density contrast $\delta_\mathrm{d}^l$, is proportional to the collapsed fraction in that region. Assuming that only halos with mass above some minimum threshold, $M_\mathrm{min}$, can form stars
\begin{equation}
\label{eq:extended_press_schecter}
x_\mathrm{HI}(\delta_\mathrm{d}^l) = 1 - \zeta \mathrm{erfc} \left( \frac{\delta_\mathrm{d}^c-\delta_\mathrm{d}^l}{\sqrt[]{2(\sigma^2_\mathrm{min}-\sigma^2_l)}} \right) \,,
\end{equation}
where $\delta_\mathrm{d}^c$ is the critical density for collapse in the spherical collapse model, and $\sigma^2_\mathrm{min}$ and $\sigma^2_\mathrm{l}$ are the variances of linear density fluctuations on the scales $R_\mathrm{min} = \left( 3 M_\mathrm{min}/4 \pi \bar{\rho}\right)^{1/3}$ and $l$ respectively. The proportionality factor $\zeta$ encodes the efficiency of halos to produce and deliver ionizing photons to the IGM, as well as the losses due to recombinations. For comparison of this model against our numerical results, we calibrate the reionization history of the model by adjusting $\zeta$ to match the reionization history of the FN-600 simulation.  Noting that $\bar{x}_\mathrm{HI}=x_\mathrm{HI}(\delta_\mathrm{d}^l)$, we can differentiate eq.~(\ref{eq:extended_press_schecter}) to compute the bias parameters. As can be seen in eq.~(\ref{eq:f_21d}) we also need an expression for the derivative, $\mathrm{d} b_1/\mathrm{d}\delta_\mathrm{d}^l$. Note that this can be written in terms of $b_1$ and $b_2$ through
\begin{equation}
\frac{d b_1}{d\delta_\mathrm{d}^l} \Big|_{\delta_\mathrm{d}^l=0} = -\left( \frac{1}{\bar{x}_\mathrm{HI}} \frac{d\bar{x}_\mathrm{HI}}{d\delta} \right)^2 + \frac{1}{\bar{x}_\mathrm{HI}} \frac{d^2\bar{x}_\mathrm{HI}}{d \delta^2} = -b^2_1+b_2\,.
\end{equation}
Inserting these relations in eq.~(\ref{eq:f_21d}), we thus obtain eq. (\ref{eq:f_21d_final}) for our first order estimate of the 21cm-matter response.  This can be easily translated to the 21cm-21cm response function, which is given in eq. (\ref{eq:f_2121_final}).

\begin{figure}[t]
 \centering
  \includegraphics[width=\textwidth]{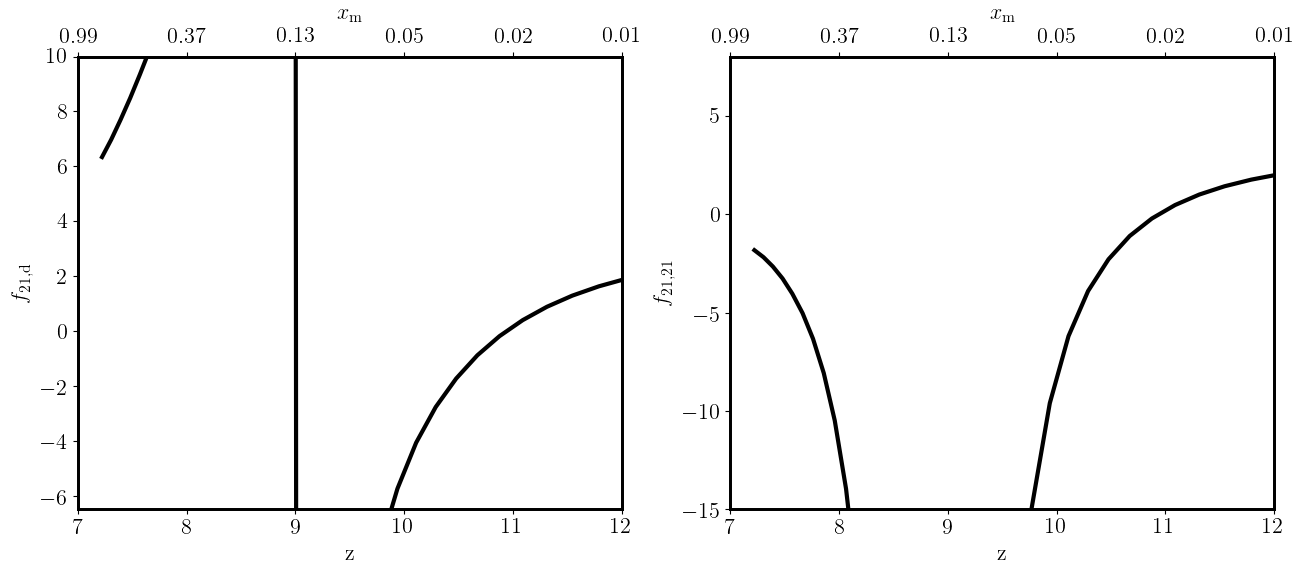}
 \caption{Redshift evolution of the response functions at each scale predicted by the analytical model. The 21cm-21cm-matter and 21cm-21cm-21cm response functions are given in left and right panel respectively. The model falls into mathematical singularity when $b_1 \rightarrow -1$. Until this time, both the response functions have similar evolution.}
 \label{fig:response_vs_z_toy}
 \end{figure}

According to eq.~\ref{eq:f_21d_final} the redshift evolution of $f_\mathrm{21,d}$ only depends on the evolution of $b_1$ and $b_2$. Figure~\ref{fig:response_vs_z_toy} shows the
results for $f_\mathrm{21,d}$ and $f_\mathrm{21,21}$ using an estimate of ${f}_\mathrm{d,d}$ for $k\approx 0.5$~Mpc$^{-1}$ taken from our simulation results at $z\approx 12$. Our model predicts that there will be a time when $f_\mathrm{21,d}$ will change sign.  This occurs when the (linear) fluctuations in the neutral hydrogen density are perfectly anti-correlated with the matter density fluctuations such that $b_1 = -1$.  Both ${f}_\mathrm{21,d}$ and ${f}_\mathrm{21,21}$ display a similar evolution until this epoch. Afterwards, ${f}_\mathrm{21,d}$ is positive and decreases from a high value whereas ${f}_\mathrm{21,21}$ is negative and increases from a low value.  In Section \ref{sec:mainresults} we find that this simple model is qualitatively similar to the numerical results from our radiative transfer simulation of reionization.

 \section{Back-of-the-envelope estimation of error on the response function}
\label{sec:back-of-envelope}
We expect to measure the bispectrum of 21-cm signal during reionization using the future radio  observations of the SKA. The response function ($\hat{f}_\mathrm{21,21}$) of large-scale fluctuations to the small-scale ones will provide us the estimate of the squeezed-limit bispectrum. However, the calculated response function will be affected by the instrumental limitations. In this section, we provide a back-of-the-envelope estimation of the error on the $\hat{f}_\mathrm{21,21}(k)$ constructed from radio observations and the sample variance.

Based on ref~\cite{chiangthesis:2015}, we give the variance on the estimated integrated bispectrum $\mathrm{var}[\hat{iB}(k)]$ as follows,
\begin{eqnarray}
    \mathrm{var}[\hat{iB}(k)] \approx \frac{V_L}{V_r N_{kL}}\Bar{\sigma}^2_{L}P_L^2(k) \ , 
\end{eqnarray}
\noindent where $V_L$, $V_r$ and $N_{kL}$ are the volume of the subvolume, volume of the entire box and the number of Fourier modes in the each $k$-bin considered while calculating the power spectra of the subvolumes. The quantities $\Bar{\sigma}^2_{L}$ and $\Bar{P}_L(k)$ are the average variance and power spectrum of all the subvolumes (defined in Section~2.3) respectively. 
Assuming the thermal noise of the telescope to be additive offset to the signal, we get the total variance on $\hat{iB}(k)$ as
\begin{eqnarray}
    \mathrm{var}[\hat{iB}(k)] \approx \frac{V_L}{V_r N_{kL}}(\Bar{\sigma}^2_{L}+\Bar{\sigma}^2_\mathrm{noise})(P_L^2(k)+P_\mathrm{noise}^2(k)) \ , 
\end{eqnarray}
\noindent where $\sigma_\mathrm{noise}$ and $P_\mathrm{noise}$ are the rms of the image constructed from the radio observations and the power spectrum of the thermal noise.


\begin{table*}
	\centering
	\caption{Table gives the telescope configuration considered in this work. We take the telescope parameters for SKA-Low taken from the latest configuration document: \url{http://astronomers.skatelescope.org/documents/}.}
	\label{tab:tele_param}
	\begin{tabular}{lcc} 
		\hline
		Parameters  & Values  \\
		\hline
		Observation time ($t_\mathrm{int}$)  & 1000 h \\
        System temperature & $60\left(\frac{300~\mathrm{MHz}}{\nu}\right)^{2.55}$ \\
        Effective collecting area ($A_\mathrm{eff}$) & $962~\mathrm{m}^2$  \\
        Core area ($A_\mathrm{core}$) & $785000~\mathrm{m}^2$ \\
        Bandwidth ($\mathcal{B}$) & 10 MHz \\
        Number of stations ($N_\mathrm{stat}$) & 224
		\\ \hline
	\end{tabular}
\end{table*}

We consider a simple estimate of the noise power spectrum $P_\mathrm{noise}(k)$ based on ref.~\cite{2013ExA....36..235M}, which is given as:
\begin{eqnarray}
    P_\mathrm{noise}(k) = \frac{4\pi k^{-3/2}}{\Delta \mathrm{ln}k}(D^2_\mathrm{c}\Delta D_\mathrm{c} \Omega_\mathrm{FOV})^{1/2}\frac{T^2_\mathrm{sys}}{\mathcal{B}~t_\mathrm{int}}\frac{A_\mathrm{core}}{N^2_\mathrm{stat}A_\mathrm{eff}}\ ,
\end{eqnarray}
\noindent where $D_\mathrm{c}$ and $\Delta D_\mathrm{c}$ are the comoving distance to the observed redshift $z$ and the comoving distance corresponding to the bandwidth $\mathcal{B}$ respectively. The solid angle subtended by the field of view $\Omega_\mathrm{FOV}$ is $\frac{\lambda^2_\mathrm{21}}{A_\mathrm{eff}}$, where $\lambda_\mathrm{21}$ and $A_\mathrm{eff}$ are the observed wavelength of 21-cm signal and the effective collecting area of the antennae station respectively. The $A_\mathrm{core}$, $T_\mathrm{sys}$ and $t_\mathrm{int}$ represent the core area of the telescope, the system temperature of the telescope and total observation time respectively. The above estimate for $P_\mathrm{noise}(k)$ considers logarithm binning of $k$ values and the bin-width is given with $\Delta \mathrm{ln}k$.
We consider the telescope parameters for the low frequency band of SKA (SKA-Low), which is shown in Table~\ref{tab:tele_param}.
The $\sigma_\mathrm{noise}$ for a fully covered $uv$ space can be given as \cite[e.g.][]{2007MNRAS.382..809D,giri2018optimal},
\begin{eqnarray}
    \sigma_\mathrm{noise} = \frac{2k_BT_\mathrm{sys}}{A_\mathrm{eff}\sqrt{\mathcal{B}~t_\mathrm{int}~N_\mathrm{stat}(N_\mathrm{stat}-1)}} \ ,
\end{eqnarray}
where $k_B$ is the Boltzmann constant. 

 \begin{figure}[t] 
 \centering
  \includegraphics[width=1.0\textwidth]{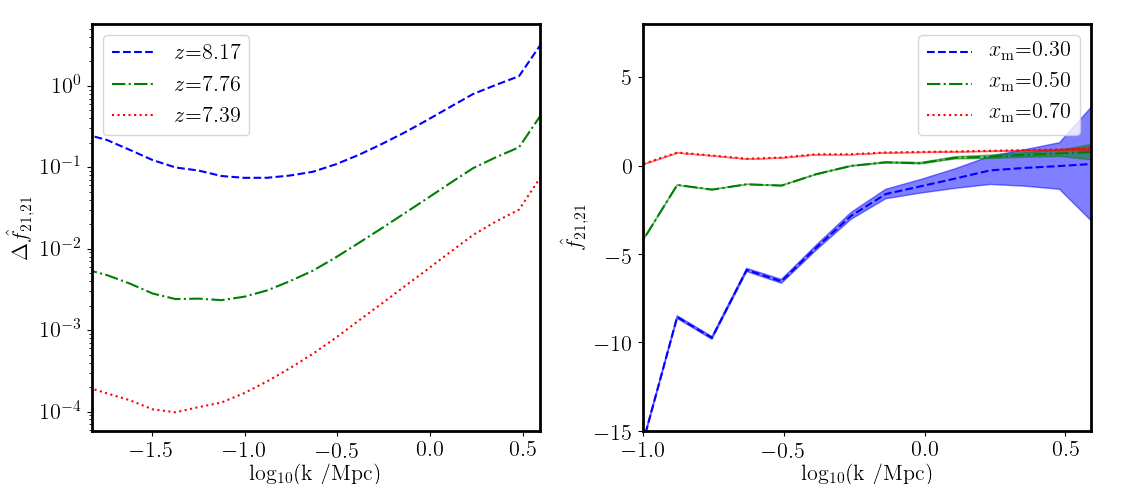}
 \caption{The error on the response function due to sample variance and thermal noise in the radio observations of 1000 h. The left panel shows that the error at three different redshifts (marked in the legend). The
 The curves in the right panel give the response function $\hat{f}_\mathrm{21,21}$ at the same redshifts when the Universe is 30\%, 50\% and 70\% ionized. The shaded region gives the 1$\sigma$ error on the $\hat{f}_\mathrm{21,21}$.}
 	\label{fig:error_on_response}
 \end{figure}

Assuming that the variance is dominated by the bispectrum term instead of the normalization, we can get the variance on the response function ($\mathrm{var} [\hat{f}]$) by dividing $\mathrm{var}[\hat{iB}(k)]$ by $P_L^2(k)$ and $\Bar{\sigma}^4_{L}$.
For a detailed calculation see appendix B of ref~\cite{chiangthesis:2015}. The error on our estimated response function $\Delta \hat{f}_\mathrm{21,21}$ will be $\sqrt{\mathrm{var} [\hat{f}]}$
In the left panel of Figure~\ref{fig:error_on_response}, we show the error estimated for the response function at $z=8.17,7.76,7.39$ ($x_\mathrm{m}=0.3, 0.5, 0.7$). While the sample variance dominates the error at large scales, the thermal noise dominates at the small scales. In the right panel of Figure~\ref{fig:error_on_response}, we show the response functions along with the error with the shaded region.
The response function constructed from the observations will be reliable to distinguish between source models when 0.03 Mpc$^{-1} < k <$ 3 Mpc$^{-1}$. 
We should note that our error calculation is too simplistic. We will investigate it further in the future.
 
%


\bibliographystyle{JHEP}
\bibliography{references2017}







\end{document}